\documentclass[aps,prd,groupedaddress,showpacs,superscriptaddress]{revtex4}
\usepackage{graphicx}
\usepackage{dcolumn}
\usepackage{bm}

\begin{document}

\title{Lightest Scalar Resonances and the Dynamics of the
\boldmath $\gamma\gamma\to\pi\pi$ Reactions}

\author{N.N. Achasov}
\email[]{achasov@math.nsc.ru}
\author{G.N. Shestakov}
\email[]{shestako@math.nsc.ru}

\affiliation{Laboratory of Theoretical Physics,
S.L. Sobolev Institute for Mathematics, 630090, Novosibirsk, Russia}

\date{\today}

\begin{abstract}
The high-statistics Belle data on the
$\gamma\gamma$\,$\to$\,$\pi^+\pi^-$ and
$\gamma\gamma$\,$\to$\,$\pi^0\pi^0$ reactions have been jointly
analyzed. The main dynamical mechanisms of these reactions for
energies below 1.5 GeV have been revealed. It has been shown that
the direct coupling constants of the $\sigma(600)$ and $f_0(980)$
resonances with a $\gamma\gamma$ pair are small and that the
$\sigma(600)$\,$\to$\,$\gamma\gamma$ and
$f_0(980)$\,$\to$\,$\gamma\gamma$ decays are four-quark transitions
due primarily to $\pi^+\pi^-$ and $K^+K^-$ loop mechanisms,
respectively. The role of the chiral shielding of the $\sigma(600)$
resonance is emphasized. The widths of the
$f_0(980)$\,$\to$\,$\gamma\gamma$ and $\sigma(600)$\,$\to$\,$
\gamma\gamma$ decays averaged over the resonance mass distributions,
as well as the width of the $f_2(1270)$\,$\to$\,$\gamma\gamma$
decay, are estimated as $\langle\Gamma_{f_0\to\gamma
\gamma}\rangle_{\pi\pi}\approx 0.19$ keV, $\langle\Gamma_{\sigma
\to\gamma\gamma}\rangle_{\pi\pi}\approx 0.45$ keV, and
$\Gamma_{f_2\to\gamma\gamma}(m^2_{f_2})\approx 3.8$ keV.
\end{abstract}
 \maketitle

The investigation of the lightest scalar resonances $\sigma(600)$,
$\kappa(800)$, $a_0(980)$, and $f_0(980)$ is one of the main goals
of nonperturbative QCD, because the elucidation of their nature is
important for understanding both the physics of confinement and the
means of the breaking of the chiral symmetry at low energies, which
are the main consequences of QCD for hadron physics. The nontrivial
nature of these states is commonly accepted. In particular, there is
plenty of evidence of their four-quark $(q^2\bar q^2)$ structure
(see, e.g., [1] and references therein). One of these evidences is
the suppression of the production of the $a_0(980)$ and $f_0(980)$
resonances in the $\gamma\gamma$\,$\to$\,$\pi^0\eta$ and $\gamma
\gamma$\,$\to$\,$\pi\pi$ reactions, respectively, which was
predicted more than 25 years ago [2] and observed in the experiment
[3]. The problem of the mechanisms of the production of the
$\sigma(600)$, $f_0(980)$, and $a_0(980)$ resonances in the
$\gamma\gamma$ collisions is closely associated with the problem of
their internal quark structure. This explains the long-term
theoretical and experimental interest in the $\gamma\gamma$\,$\to
$\,$\pi\pi$ reactions at low energies. Recently, the Belle
Collaboration obtained new data on the cross sections for the
$\gamma\gamma\to\pi^+\pi^-$ [4] and
$\gamma\gamma$\,$\to$\,$\pi^0\pi^0$ [5] reactions with statistics
two orders of magnitude larger than all previous experiments and
revealed a pronounced signal from the $f_0(980)$ resonance [4,5].
The preceding indications of the production of the $f_0(980)$
resonance in the $\gamma\gamma$ collisions were much less definite
[6--8]. The signal from the $f_0(980)$ resonance appears to be
small, which is in good agreement with the prediction of the
four-quark model [1,2].

In this paper, we report the results of the investigation of the
main dynamical mechanisms of the $\gamma\gamma$\,$\to$\,$\pi^+\pi^-$
and $\gamma\gamma$\,$\to$\,$\pi^0\pi^0$ reactions on the basis of
the analysis of the Belle data [4,5] and our previous investigations
of the physics of the scalar mesons in the $\gamma\gamma$ collisions
[2,9--13].

The Belle data on the cross sections for the $\gamma\gamma$\,$\to
$\,$\pi^+\pi^-$ and $\gamma\gamma$\,$\to$\,$\pi^0\pi^0$ reactions
obtained for invariant mass $\sqrt{s}$ of the $\pi\pi$ systems from
0.8 to 1.5\,GeV are shown in Fig. 1, where the data of other groups
[6--8] are also shown for $\sqrt{s}$ from $2m_\pi$ to 0.85\,GeV. All
existing data correspond to the incomplete solid angle of the
detection of the final pions such that $|\cos \theta|\leq0.6$ and
$|\cos \theta|\leq0.8$ for the production of the $\pi^+\pi^-$ and
$\pi^0\pi^0$ pairs, respectively, where $\theta$ is the polar angle
of the pion emission in the cms of the initial photons. The
pronounced peaks attributed to the production of the $f_0(980)$ and
$f_2(1270)$ resonances are observed in the cross sections for both
reactions. The background under these peaks is nearly absent in the
$\gamma\gamma$\,$\to$\,$\pi^0\pi^0$ channel. On the contrary, the
resonances in the $\gamma\gamma$\,$\to$\,$\pi^+\pi^-$ channel are
seen against a large smooth background, which is primarily
attributed to the mechanism of the charged one-pion exchange
[11--16]. The pure Born cross section for the
$\gamma\gamma$\,$\to$\,$\pi^+\pi^-$ process at $|\cos
\theta|\leq0.6$, the total cross section
$\sigma^{\mbox{\scriptsize{Born}}}$\,=\,$\sigma^{\mbox
{\scriptsize{Born}}}_0$\,+\,$\sigma^{\mbox {\scriptsize{Born}}}_2$,
and the cross sections $\sigma^{\mbox{\scriptsize{Born}}}_0$ and
$\sigma^{\mbox{\scriptsize{Born}}}_2$, where the subscript
($\lambda$\,=\,$0$ or 2) is the absolute value of the difference
between the helicities of the initial photons, are shown in Fig. 1a
for comparison. Owing to the Low theorem and chiral symmetry, the
one-pion Born contribution should dominate near the threshold of the
$\gamma\gamma$\,$\to$\,$\pi^+\pi^-$ reaction. As seen in Fig. 1a,
this expectation does not contradict the near-threshold data;
however, these data were obtained with large errors. The cross
section $\sigma^{\mbox{\scriptsize{Born}}}_0$ decreases rapidly with
an increase in $\sqrt{s}$, so that the contribution
$\sigma^{\mbox{\scriptsize{Born}}}_2$ dominates completely in
$\sigma^{\mbox{\scriptsize{Born}}}$  at $\sqrt{s}>$\,0.5\,GeV (see
Fig. 1a). Note that the contributions from the $S$ and $D_{\lambda
=2}$ partial waves dominate in the region $\sqrt{s}<$\,1.5\,GeV in
$\sigma^{\mbox{\scriptsize {Born}}}_0$ and
$\sigma^{\mbox{\scriptsize{Born}}}_2$, respectively. These partial
Born contributions are strongly modified due to the strong
interaction between pions in the final state, because the $\pi\pi$
interaction at $\sqrt{s}<$\,1.5\,GeV is strong only in the $S$ and
$D$ waves. The inclusion of the final-state interaction in the
$S$-wave Born amplitudes of the $\gamma\gamma$\,$\to$\,$\pi^+\pi^-$
(and $\gamma\gamma$\,$\to$\,$K^+K^-$) reaction leads to certain
predictions for the $S$-wave amplitude of the $\gamma\gamma$\,$\to
$\,$\pi^0\pi^0$ reaction.

\begin{figure} \includegraphics{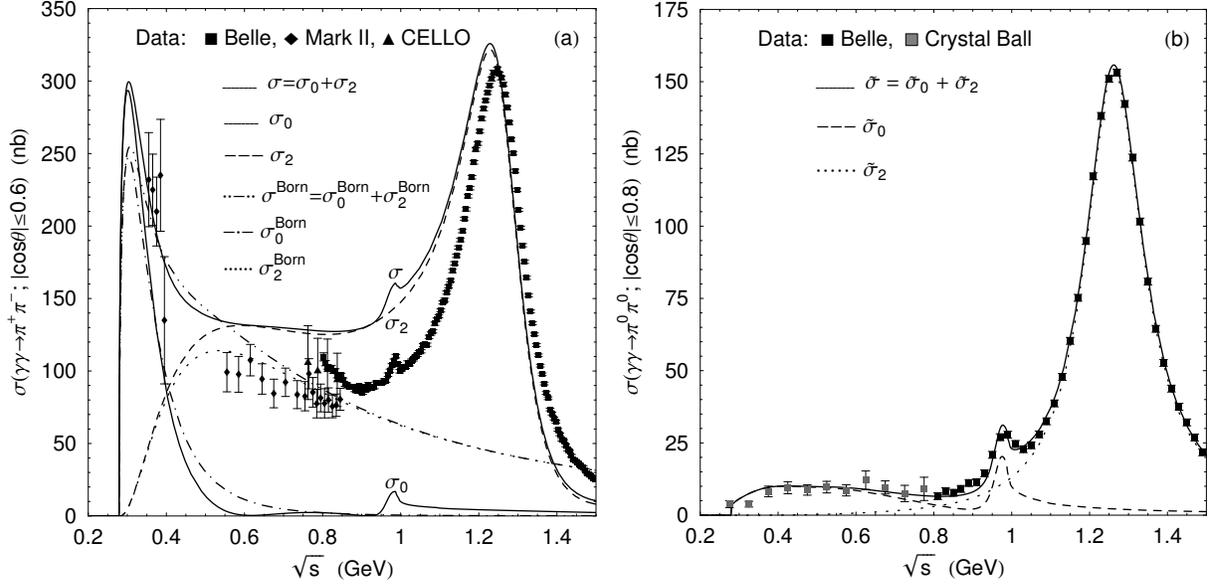} \caption {Cross
sections for the $\gamma\gamma\to\pi^+\pi^-$ and $\gamma\gamma\to
\pi^0\pi^0$ reactions. Only statistical errors are shown for the
Belle data [4,5]. The curves in panel (a) are described in the main
text and on the figure. The curves in panel (b) are the
approximations of the data on the $\gamma\gamma\to\pi^0\pi^0$
reaction.}\end{figure}

\begin{figure} \includegraphics{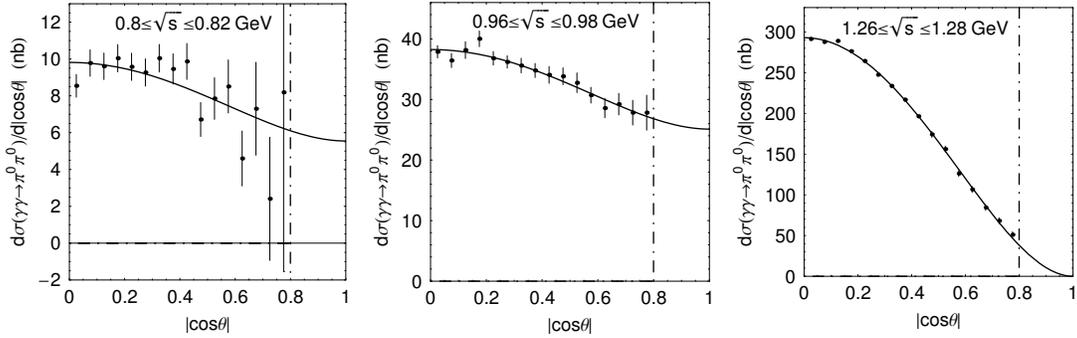} \caption
{Angular distributions in the $\gamma\gamma\to \pi^0\pi^0$ reaction.
The Belle experimental data are taken from [5]. The vertical
straight line $|\cos\theta|$\,=\,0.8 is the boundary of the region
available for the measurements. The solid lines are the
approximations.}\end{figure}

Figure 2 shows the Belle experimental data for the angular
distributions in the $\gamma\gamma\to\pi^0\pi^0$ reaction [5]. They
are excellently reproduced by the simple two-parametric expression
$|a|^2+|b\,d^2_{20}(\theta) |^2$, where $d_{\lambda0}^l(\theta)$ is
the $d$ function [3] and $l$ is the orbital angular momentum of the
final $\pi\pi$ system. Therefore, the cross section for the
$\gamma\gamma$\,$\to$\,$\pi^0\pi^0$ at $\sqrt{s}<$\,1.5\,GeV is
described by contributions only from the $S$ and $D_2$ partial waves
[17].

Thus, let us consider a model for the helicity, $M_\lambda$, and
partial, $M_{\lambda l}$, amplitudes of the $\gamma\gamma$\,$\to
$\,$\pi\pi$ reaction, where the electromagnetic Born contributions
from point-like charged $\pi$ and $K$ exchanges modified in the $S$
and $D_2 $ waves by strong final-state interactions, as well as the
contributions due to the direct interaction of the resonances with
photons (see also [11,13]), are taken into account:
\begin{eqnarray}
M_0(\gamma\gamma\to\pi^+\pi^-;s,\theta)
=M^{\mbox{\scriptsize{Born}}}_0(s,\theta)+\widetilde{
I}_{\pi^+\pi^-}(s)\,T_{\pi^+\pi^-\to\pi^+\pi^-}(s)
+\widetilde{I}_{K^+K^-}(s)\,T_{K^+K^-\to\pi^+\pi^-}(s)+M^{\mbox
{\scriptsize{direct}}}_{\mbox{\scriptsize{res}}}(s)\,,
\end{eqnarray}
\begin{eqnarray}
M_2(\gamma\gamma\to\pi^+\pi^-;s,\theta)
=M^{\mbox{\scriptsize{Born}}}_2(s,\theta)+80\pi d^2_{20}(\theta)
M_{\gamma\gamma\to f_2(1270)\to\pi^+\pi^-}(s),&
\end{eqnarray}
\begin{eqnarray}&M_0(\gamma\gamma\to\pi^0\pi^0;s,\theta)
=M_{00}(\gamma\gamma\to\pi^0\pi^0;s)& \nonumber\\ & =
\widetilde{I}_{\pi^+\pi^-}(s)\,T_{\pi^+\pi^-\to\pi^0\pi^0}(s)
+\widetilde{I}_{K^+K^-}(s)\,T_{K^+K^-
\to\pi^0\pi^0}(s)+M^{\mbox{\scriptsize{direct}}}_{\mbox{\scriptsize{res}}}
(s)\,, & \end{eqnarray}
\begin{eqnarray}
M_2(\gamma\gamma\to\pi^0\pi^0;s,\theta)
=5d^2_{20}(\theta)M_{22}(\gamma \gamma\to\pi^0\pi^0;s)=80\pi
d^2_{20}(\theta)M_{\gamma\gamma\to f_2(1270)\to\pi^0\pi^0}(s)\,.
\end{eqnarray} Here,
$M_0^{\mbox{\scriptsize{Born}}}(s,\theta)$\,=\,$(32\pi\alpha/s)/[1-
\rho^2_{\pi^+}(s)\,\cos^2\theta]$ and
$M_2^{\mbox{\scriptsize{Born}}}(s,\theta)$\,=\,$8\pi\alpha\,\rho^2_{\pi^+}
(s)\,\sin^2\theta/[1-\rho^2_{\pi^+}(s)\,\cos^2\theta]$ are the Born
helicity amplitudes of the $\gamma\gamma\to\pi^+\pi^-$ reaction,
$\rho_{\pi^+}(s)$\,=\,$(1-4m^2_{\pi^+}/s)^{1/2}$, and
$\alpha$\,=\,1/137. The function $\widetilde{I}_{\pi^+\pi^-}(s)$ at
$s\geq4m^2_{\pi^+}$ has the form
$\widetilde{I}_{\pi^+\pi^-}(s)=8\alpha \left\{
\frac{m^2_{\pi^+}}{s}\left[\pi+i\ln\frac{1+\rho_{\pi^+}(s)}
{1-\rho_{\pi^+}(s)}\right]^2-1\right\},$
$\mbox{Im}\widetilde{I}_{\pi^+\pi^-}(s)$\,=\,$\rho_{\pi^+}(s)M_{00}^{
\mbox{\scriptsize{Born}}}(s)$, and $\widetilde{I}_{K^+K^-}(s)$ at
$s\geq4m^2_{K^+}$ is obtained from $\widetilde{I}_{\pi^+\pi^-}(s)$
by changing $m_{\pi^+}$ to $m_{K^+}$ and $\rho_{\pi^+}(s)$ to
$\rho_{K^+}(s)$\,=\,$(1-4m^2_{K^+}/s)^{1/2}$;
$\rho_{K^+}(s)$\,$\to$\,$i|\rho_{K^+}(s)|$ if 0\,$<s<$\,$
4m^2_{K^+}$. The functions $\widetilde{I}_{\pi^+\pi^-}(s)$ and
$\widetilde{I}_{K^+K^-}(s)$ are the amplitudes of the triangle
diagrams $\gamma\gamma$\,$\to$\,$\pi^+\pi^-$\,$\to$\,$\sigma$,\,$
f_0$ and $\gamma\gamma$\,$\to$\,$K^+K^-$\,$\to$\,$\sigma$,\,$f_0$
(and other scalar resonances); $T_{\pi^+\pi^-\to\pi^+\pi^-}(s)$,
$T_{\pi^+\pi^-\to\pi^0\pi^0}(s)$,
$T_{K^+K^-\to\pi^+\pi^-}(s)$\,=\,$T_{K^+K^-\to\pi^0\pi^0}(s)
$\,=\,$T_{\pi^+\pi^-\to K^+K^-}(s)$ are the $S$wave amplitudes of
the corresponding reactions; $T_{
\pi^+\pi^-\to\pi^+\pi^-}(s)$\,=\,$[2T^0_0(s)+T^2_0(s)]/3$ and
$T_{\pi^+\pi^-\to\pi^0\pi^0}(s)$\,=\,$2[T^0_0(s)-T^2_0(s)]/3$, where
$T^I_0(s)$\,=\,$\{\eta^I_0(s)\exp[2i\delta^I_0(s)]-1\}/[2i\rho_{\pi^+}(s)]$
are the amplitudes, $\delta^I_0(s)$ are the phases, and
$\eta^I_0(s)$ are the inelasticity factors of the $S$ wave $\pi\pi$
scattering in the channels with isospin $I$\,=\,0 and 2. Really,
$\eta^0_0(s)$\,=\,1 up to the threshold of the $K\bar K$ channel.
For this reason, $T_{\pi^+\pi^-\to K^+K^-
}(s)=e^{i\delta^0_0(s)}|T_{\pi^+\pi^-\to K^+K^- }(s)|$ at
$4m_\pi^2$\,$<s<$\,$4m^2_K$ [11,13,18]. We also set
$\eta^2_0(s)$\,=\,1 at all $s$ values under consideration and take
$\delta^2_0(s)$ from [19]. Expressions (1) and (3) imply that the
amplitudes $T_{\pi^+\pi^-\to\pi\pi}(s)$ and $T_{K^+K^-\to\pi\pi}(s)$
in $\gamma\gamma$\,$\to$\,$\pi^+\pi^-$\,$\to$\,$\pi\pi$ and
$\gamma\gamma$\,$\to$\,$K^+K^-$\,$\to$\,$\pi\pi$ rescattering loops
are on-mass-shell amplitudes. Note that the unitarity condition are
satisfied in the model under consideration [13].

The parametrization of the amplitudes  $T^0_0(s)$ and
$T_{K^+K^-\to\pi^0\pi^0}(s)$, which was used in the joint analysis
of the data on the $\pi^0\pi^0$ mass spectrum in the
$\phi$\,$\to$\,$\pi^0\pi^0\gamma$ decay, $\pi\pi$ scattering at
$2m_\pi<$\,$\sqrt{s}<$\,1.6\,GeV, and $\pi\pi$\,$\to$\,$K\bar K$
reaction, was described in detail in [18]. This parametrization is
based on the concept that the amplitude $T^0_0(s)$ must include the
contribution from mixed $\sigma(600)$ and $f_0(980)$ resonances and
the contribution from the background, which has a large negative
phase due to the chiral symmetry; the latter contribution shields
(hides) the $\sigma(600)$ resonance [1,18,20]. Formulas (1) and (3)
transfer the effect of the chiral shielding of the $\sigma(600)$
resonance from the $\pi\pi$ scattering to the $\gamma\gamma$\,$\to
$\,$\pi\pi$ amplitudes. If this shielding were absent, then the
$\gamma\gamma$\,$\to$\,$\pi^0\pi^0$ cross section (see Fig. 1b)
would be about 100\,nb rather than 10\,nb due to the $\pi^+\pi^-$
loop mechanism of the $\sigma(600)$\,$\to$\,$\gamma\gamma $ decay
[12]. According [18],\begin{center}
$T^0_0(s)=T^{\pi\pi}_B(s)+e^{2i\delta^{\pi\pi}_{B}(s)}
T^{\pi\pi}_{\mbox{\scriptsize{res}}}(s)\,,$\end{center}
\begin{center}$T_{K^+K^-\to
\pi^+\pi^-}(s)=e^{i[\delta^{\pi\pi}_{B}(s)+\delta^{K\bar K}_{B}(s)]}
T^{K\bar K\to\pi\pi}_{\mbox{\scriptsize{res}}}(s)$\end{center} and
\begin{center}$T^{\pi\pi}_B(s)=\{\exp[2i\delta^{\pi\pi}_{B}(s)]
-1\}/[2i\rho_{\pi^+}(s)]\,,$\end{center} where
$\delta^{\pi\pi}_{B}(s)$ and $\delta^{K\bar K}_{B}(s)$ are the
phases of the elastic $S$-wave background in the $\pi\pi$ and $K\bar
K$ channels with $I$\,=\,0, respectively. The amplitudes of the
$\sigma(600)-f_0(980)$ resonance complex have the form [13,18]
\begin{equation}T^{\pi\pi}_{\mbox{\scriptsize{res}}}(s)=
(\eta^0_0(s)\exp[2i\delta_{\mbox{\scriptsize{res}}}(s)]-1)/
[2i\rho_{\pi^+}(s)]=3\,\frac{g_{\sigma\pi^+\pi^-}\Delta_{f_0}(s)+
g_{f_0\pi^+\pi^-}\Delta_\sigma(s)}
{32\pi[D_\sigma(s)D_{f_0}(s)-\Pi^2_{f_0\sigma}(s)]}\,,
\end{equation}
\begin{equation}
T^{K\bar K\to\pi\pi}_{\mbox{\scriptsize{res}}}(s)=\frac{g_{\sigma
K^+K^-}\Delta_{f_0}(s)+g_{f_0K^+K^-}\Delta_\sigma(s)}
{16\pi[D_\sigma(s)D_{f_0}(s)-\Pi^2_{f_0\sigma}(s)]}\,, \\
\end{equation}
\begin{equation}M^{\mbox{\scriptsize{direct}}}_{\mbox{\scriptsize{res}}}(s)=
s\,e^{i\delta^{\pi\pi}_B(s)}\,
\frac{g^{(0)}_{\sigma\gamma\gamma}\Delta_{f_0}(s)+
g^{(0)}_{f_0\gamma\gamma}\Delta_\sigma(s)}
{D_\sigma(s)D_{f_0}(s)-\Pi^2_{f_0\sigma}(s)}\,,
\end{equation} where
$\Delta_{f_0}(s)$\,=\,$D_{f_0}(s)g_{\sigma\pi^+\pi^-}+\Pi_{f_0\sigma}
(s)g_{f_0\pi^+\pi^-}$,
$\Delta_\sigma(s)=D_\sigma(s)g_{f_0\pi^+\pi^-}+\Pi_{f_0\sigma}
(s)g_{\sigma\pi^+\pi^-}$, and $\delta^0_0(s)=\delta^{\pi\pi}_B(s)+
\delta_{\mbox{\scriptsize{res}}}(s)$. The expressions presented in
[18] were used for $\delta^{\pi\pi}_B(s)$, propagators
$1/D_\sigma(s)$ and $1/D_{f_0}(s)$ of the $\sigma(600)$ and
$f_0(980) $ resonances, respectively, and the matrix element of the
polarization operator $\Pi_{f_0\sigma}(s)$. The values of the
parameters in the strong amplitudes ($m_\sigma$,
$g_{\sigma\pi^+\pi^-}$, $g_{f_0K^+K^-}$, etc.) correspond to variant
1 from Table 1 in [18].

Thus, according to Eqs. (1), (3), and (7), the $\sigma(600)$\,$\to
$\,$\gamma\gamma$ and $f_0(980)$\,$\to$\,$\gamma\gamma$ decays are
described by the triangle $\pi^+\pi^-$ and $K^+K^-$ loop diagrams
(the resonances\,$\to$\,$\pi^+\pi^-$,
$K^+K^-$\,$\to$\,$\gamma\gamma$), which correspond to the four-quark
transitions [12,13], and by the direct coupling constants of the
resonances with the photons $g^{(0)}_{\sigma\gamma\gamma}$ and
$g^{(0)}_{f_0\gamma\gamma}$ [9--14].

The amplitudes of the production of the $f_2(1270)$ resonance in
Eqs. (2) and (4), $M_{\gamma\gamma\to
f_2(1270)\to\pi^+\pi^-}(s)=M_{\gamma\gamma\to
f_2(1270)\to\pi^0\pi^0}(s)$, have the form \begin{center}
$\sqrt{s}\,G_2(s)
\sqrt{2\Gamma_{f_2\to\pi\pi}(s)/3}\Bigm/[m^2_{f_2}-s-i
\sqrt{s}\Gamma^{\mbox{\scriptsize{tot}}}_{f_2}(s)]\,,$\end{center}
where
$$G_2(s)=\sqrt{\Gamma^{(0)}_{f_2\to\gamma\gamma}(s)}+i\frac{M_{22}^{
\mbox{\scriptsize{Born}}}(s)}{16\pi}
\sqrt{\frac{2}{3}\rho_{\pi^+}(s)\Gamma_{f_2\to\pi\pi}(s)}\,,$$
\begin{center}
$\Gamma^{\mbox{\scriptsize{tot}} }_{f_2}(s)
$\,=\,$\Gamma_{f_2\to\pi\pi}(s)+\Gamma_{f_2\to K\bar
K}(s)+\Gamma_{f_2\to4\pi}(s)\,.$\end{center} By definition
$$\Gamma_{f_2\to\gamma\gamma}
(s)=|G_2(s)|^2\qquad\mbox{and}\qquad
\Gamma^{(0)}_{f_2\to\gamma\gamma}(s)=\frac{m_{f_2}}{\sqrt{s}}
\Gamma^{(0)}_{f_2\to\gamma\gamma}(m^2_{f_2})
\frac{s^2}{m^4_{f_2}}\,.$$ Here, the factor $s^2$ and the factor $s$
in Eq. (7) appear due to the gauge invariance. The second term in
$G_2(s)$ corresponds to the $f_2(1270)$\,$\to$\,$\pi^+\pi^-
$\,$\to$\,$\gamma\gamma$ transition with real pions in the
intermediate state and ensures the satisfaction of the Watson
theorem for the $\gamma\gamma$\,$\to$\,$\pi\pi$ amplitude with
$\lambda$\,=\,$l$\,=\,2 and $I$\,=0 below the first inelastic
threshold. This term makes a small contribution (less than 6\%) to
$\Gamma_{f_2\to\gamma\gamma}(m^2_{f_2})$ [13]. It is commonly
accepted that the quark-antiquark transition $q\bar q$\,$\to$\,$
\gamma\gamma$, i.e., the
$\Gamma^{(0)}_{f_2\to\gamma\gamma}(m^2_{f_2})$ contribution
dominates in the $f_2(1270)$\,$\to$\,$\gamma\gamma$ decay. As shown
in [12,13] and noted below, the situation for the scalar mesons is
opposite.

The leading contribution to $\Gamma^{\mbox{\scriptsize{tot}}
}_{f_2}(s)$ comes from the partial decay width
$f_2(1270)$\,$\to$\,$\pi\pi$, $$
\Gamma_{f_2\to\pi\pi}(s)=\Gamma^{\mbox{\scriptsize{tot}}
}_{f_2}(m^2_{f_2})B(f_2\to\pi\pi)\frac{m^2_{f_2}}{s}
\frac{q^5_{\pi^+}(s)}{q^5_{\pi^+}(m^2_{f_2})}
\frac{D_2(q_{\pi^+}(s)R_{f_2})}{D_2(q_{\pi^+}(m^2_{f_2})R_{f_2})}\,,$$
where $D_2(x)$\,=\,$1/(9+3x^2+x^4)$,
$q_{\pi^+}(s)$\,=\,$\sqrt{s}\rho_{\pi^+}(s)/2$, $R_{f_2}$ is the
interaction radius, and $B(f_2$\,$\to$\,$\pi\pi)$\,=\,0.847. Small
contributions from $\Gamma_{f_2\to K\bar K}(s)$ and
$\Gamma_{f_2\to4\pi}(s)$ are the same as in [13]. The parameter
$R_{f_2}$ [4--8,13] controls the relative shape of the wings of the
$f_2(1270)$ resonance and is important particulary for the
approximation of the data with small errors.

We use the following notation and normalizations for the cross
sections
$$\sigma(\gamma\gamma\to\pi^+\pi^+;|\cos\theta|\leq0.6)\equiv
\sigma=\sigma_0+\sigma_2\qquad\mbox{and}\qquad\sigma(\gamma\gamma
\to\pi^0\pi^0;|\cos\theta|\leq0.8)\equiv\tilde{\sigma}=
\tilde{\sigma}_0+\tilde{\sigma}_2\,,$$ where $$\sigma_\lambda=\frac{
\rho_{\pi^+}(s)}{64\pi s}\int^{0.6}_{-0.6}|M_\lambda(\gamma
\gamma\to\pi^+\pi^+;s,\theta)|^2d\cos\theta\qquad\mbox{and}\qquad
\tilde{\sigma}_\lambda=\frac{\rho_{\pi^+}(s)}{128\pi
s}\int^{0.8}_{-0.8}|M_\lambda(\gamma\gamma\to\pi^0\pi^0;s,\theta)|^2
d\cos\theta\,.$$

First, we consider the approximation of the data only on the cross
section for the $\gamma\gamma$\,$\to$\,$\pi^0\pi^0$ reaction (see
Fig. 1b); as mentioned above, the background situation in this
channel is more pure than in the $\gamma\gamma$\,$\to$\,$\pi^+
\pi^-$ one. The solid line in Fig. 1b, which well describes these
data, corresponds to the following model parameters:
$m_{f_2}$\,=\,1.269\,GeV,
$\Gamma^{\mbox{\scriptsize{tot}}}_{f_2}(m^2_{f_2})$\,=\,0.182\,GeV,
$R_{f_2}$\,=\,8.2\,GeV$^{-1}$, $\Gamma_{f_2\to\gamma\gamma
}(m_{f_2})$\,=\,3.62\,keV, $m_{f_0}$\,=\,0.969\,GeV,
$g^{(0)}_{\sigma\gamma\gamma}$\,=\,0.536\,GeV$^{-1}$, and
$g^{(0)}_{f_0\gamma\gamma}$\,=\,0.652\,GeV$^{-1}$. The approximation
indicates that the direct constants $g^{(0)}_{\sigma\gamma\gamma}$
and $g^{(0)}_{f_0\gamma \gamma}$ are small in agreement with the
prediction in [2]: \, $\Gamma^{(0)}_{\sigma\to\gamma
\gamma}(m^2_{\sigma})$\,=\,$|m^2_\sigma
g^{(0)}_{\sigma\gamma\gamma}|^2/(16\pi m_\sigma)$\,=\,0.012\,keV and
$\Gamma^{(0)}_{f_0\to\gamma \gamma}(m^2_{f_0})$\,=\,$|m^2_{f_0}
g^{(0)}_{f_0\gamma\gamma}|^2/(16\pi m_{f_0})$\,=\,0.008\,keV. In
turn, this indicates the dominance of the $\pi^+\pi^-$ and $K^+K^-$
loop mechanisms of the coupling of $\sigma(600)$ and $f_0(980)$ with
photons. Indeed, according to estimates [11,12], the width of the
$\sigma(600)$\,$\to$\,$\pi^+\pi^-\to\gamma \gamma$ decay through the
$\pi^+\pi^-$ loop mechanism is approximately 1--1.75\,keV in the
region 0.4\,$<\sqrt{s}<$\,0.5\,GeV [12], and the width of the
$f_0(980)$\,$\to$\,$K^+K^-$\,$\to$\,$\gamma \gamma$ decay through
the $K^+K^-$ loop mechanism after averaging over the resonance mass
distribution is approximately 0.15--0.2\,keV [11].

However, such an approximation of the $\gamma\gamma$\,$\to$\,$
\pi^0\pi^0$ cross section leads to a contradiction with the data for
$\gamma\gamma$\,$\to$\,$\pi^+ \pi^-$ (see the solid line for
$\sigma$\,=\,$\sigma_0$\,+\,$ \sigma_2$ in Fig. 1a). This is
associated with a large Born contribution to $ \sigma_2$ and a
strong constructive (destructive) interference of this contribution
with the contribution from the $f_2(1270)$ resonance at
$\sqrt{s}<$\,$m_{f_2}$ ($\sqrt{s}>$\,$m_{f_2}$). Note that these
contributions are absent in $\gamma\gamma$\,$\to$\,$\pi^0\pi^0$
reaction. The problem of the joint description of the data for the
$\gamma\gamma$\,$\to$\,$\pi^+ \pi^-$ and $\gamma\gamma$\,$\to$\,$
\pi^0\pi^0$ reactions was pointed out in [13], where the solution of
this problem was proposed. The situation can be significantly
corrected by multiplying the $\gamma\gamma$\,$\to$\,$\pi^+ \pi^-$
Born amplitudes for point particles,
$M_\lambda^{\mbox{\scriptsize{Born}}}(s,\theta)$, by the common
suppressing form factor $G(t,u)$ [7,8,10,13,16,21], where $t$ and
$u$ are the normal Mandelstam variables for the
$\gamma\gamma$\,$\to$\,$\pi^+ \pi^-$ process. To demonstrate this,
we use the following expression proposed in [21]:
$$G(t,u)=\frac{1}{s}\left[\frac{m^2_{\pi^+}-t}{1-(u-m^2_{\pi^+})/x^2_1}+
\frac{m^2_{\pi^+}-u}{1-(t-m^2_{\pi^+})/x^2_1}\right]\,,$$ where
$x_1$ is the free parameter. This ansatz is acceptable in the
physical region of the $\gamma\gamma\to\pi^+\pi^-$ reaction.
Changing $m_{\pi^+}$ to $m_{K^+}$ and $x_1$ to $x_2$, we also obtain
the form factor for the Born amplitudes of the
$\gamma\gamma$\,$\to$\,$K^+ K^-$reaction. The solid lines for the
cross sections $\sigma$\,=\,$\sigma_0$\,+\,$ \sigma_2$ and
$\tilde{\sigma}$\,=\,$ \tilde{\sigma}_0+\tilde{\sigma}_2$ in Figs.
3a and 3b, respectively, demonstrate the joint approximation of the
data for the $\gamma \gamma$\,$\to$\,$\pi^+ \pi^-$ reaction in the
region 0.85\,$<\sqrt{s}<$\,1.5\,GeV and for the
$\gamma\gamma$\,$\to$\,$\pi^0\pi^0$ reaction in the region
$2m_\pi<$\,$\sqrt{s}<$\,1.5\,GeV including the form factors
modifying the Born contributions for point particles. The resulting
description is more than satisfactory, but only with the inclusion
of the total (statistical and systematic) errors in the Belle data,
which are shown in Figs. 3a and 3b in the form of shaded bands. We
\begin{figure} \includegraphics{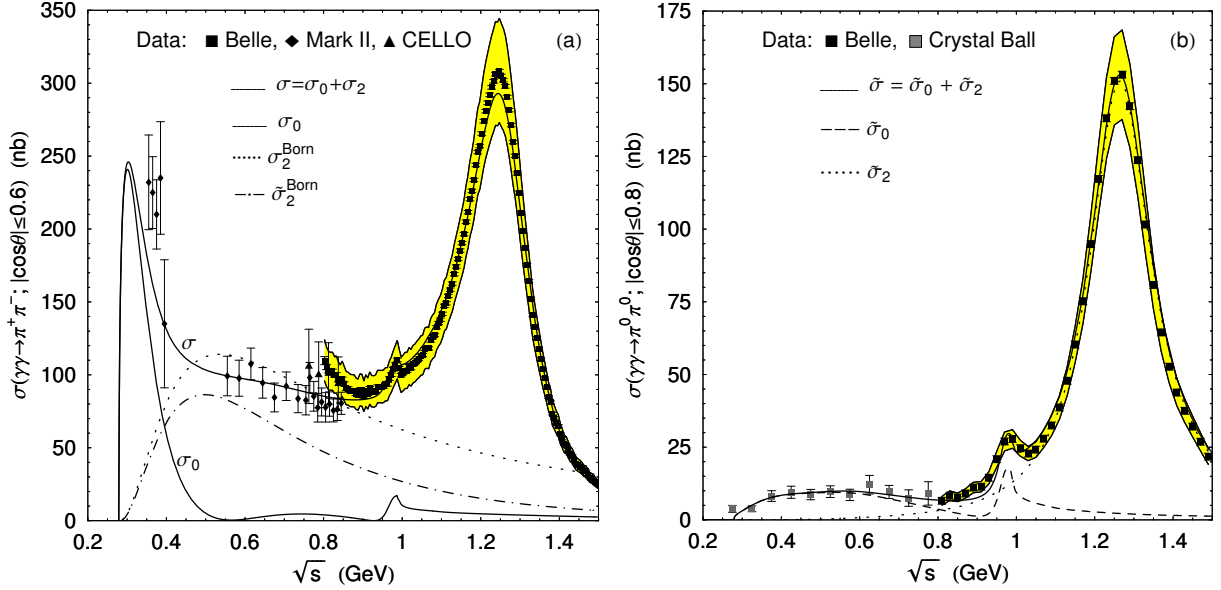} \caption
{Joint description of the data on the cross sections for the
$\gamma\gamma\to\pi^+\pi^-$ and $\gamma\gamma\to\pi^0\pi^0$
reaction. The shaded bands correspond to the Belle data [4,5] with
the statistical and systematic errors (errors are added
quadratically). The curves are described in the main text and on the
figures; $\tilde{\sigma}^{\mbox{\scriptsize{ Born}}}_2$ in panel (a)
is the Born cross section for the $\gamma\gamma\to\pi^+\pi^-$
reaction with the inclusion of the form factor.}\end{figure} believe
that this treatment is justified. The statistical errors of two
Belle measurements are small so that it is impossible to obtain the
formally acceptable $\chi^2$ values in the joint approximation of
the $\pi^+\pi^-$ and $\pi^0\pi^0$ data without the inclusion of the
systematic errors. The lines in Figs. 3a and 3b correspond to the
parameters $m_{f_2}$\,=\,1.272\,GeV,
$\Gamma^{\mbox{\scriptsize{tot}}}_{f_2}(m^2_{f_2})$\,=\,0.196\,GeV,
$R_{f_2}$\,=\,8.2\,GeV$^{-1}$, $\Gamma_{f_2\to\gamma\gamma
}(m_{f_2})$\,=\,3.84\,keV, $m_{f_0}$\,=\,0.969\,GeV,
$g^{(0)}_{\sigma\gamma\gamma}$\,=\,-0.049\,GeV$^{-1}$
($\Gamma^{(0)}_{\sigma\to\gamma \gamma}(m^2_{\sigma})$ is
negligible), $g^{(0)}_{f_0\gamma\gamma}$\,=\,0.718\,GeV$^{-1}$
($\Gamma^{(0)}_{f_0\to\gamma
\gamma}(m^2_{f_0})$\,$\approx$\,0.01\,keV), $x_1$\,=\,0.9\,GeV and
$x_2$\,=\,1.75\,GeV. A comparison of Figs. 1b and 3b shows that the
effect of the form factors on the cross section for the $\gamma
\gamma\to\pi^0\pi^0$ reaction is weak in contrast to the cross
section for the $\gamma\gamma\to\pi^+\pi^-$ (see Figs. 1a and 3a).
We emphasize that our conclusions on the mechanisms of the
two-photon decays (productions) of the $\sigma(600)$ and $f_0(980)$
resonances remain valid.

It is interesting to consider the $\gamma\gamma$\,$\to$\,$\pi^+
\pi^-$ cross section attributed only to the resonance contributions,
i.e., $$\sigma_{\mbox{\scriptsize{res}}}(\gamma\gamma\to\pi^+\pi^-
;s)= \frac{ \rho_{\pi^+}(s)}{32\pi s}|\widetilde{ I}^{\mbox{\
\scriptsize{ff}}}_{\pi^+\pi^-}(s)\,e^{2i\delta^{\pi\pi}_{B}(s)}
T^{\pi\pi}_{\mbox{\scriptsize{res}}}(s) +\widetilde{I}^{\mbox{\
\scriptsize{ff}}}_{K^+K^-}(s) T_{K^+K^-\to\pi^+\pi^-}(s)+M^{\mbox
{\scriptsize{direct}}}_{\mbox{\scriptsize{res}}}(s)|^2\,,$$ [see
Eqs. (1) and (5)--(7)]. Here, the superscript ff means that the
functions $\widetilde{I}(s)$ are obtained with the inclusion of the
form factors [10]. The cross section $\sigma_{\mbox{\scriptsize{res
}}}(\gamma\gamma$\,$\to$\,$\pi^+\pi^-;s)$ has a pronounced peak near
1 GeV from the $f_0(980)$ resonance, which is due primarily to the
contribution from the $\gamma\gamma$\,$\to$\,$K^+K^-$\,$\to$\,$
\pi^+\pi^-$ transition. Following [9,11], we determine the width of
the $f_0(980)$\,$\to$\,$\gamma\gamma$ decay averaged over the
resonance mass distribution in the $\pi\pi$ channel:
\begin{equation}\langle\Gamma_{f_0\to\gamma\gamma}\rangle_{\pi\pi}=
\int\limits_{0.8\mbox{\,\scriptsize{GeV}}}^{1.1\mbox{\,\scriptsize{GeV}}}
\frac{3s}{8\pi^2}\,\sigma_{\mbox{\scriptsize{res}}}
(\gamma\gamma\to\pi^+\pi^-;s)\,d\sqrt{s}\,.\end{equation} This
quantity is an adequate characteristic of the coupling of the
$f_0(980)$ resonance with a $\gamma\gamma$ pair [11]. For the
present joint approximation, $\langle\Gamma_{f_0\to\gamma\gamma}
\rangle_{\pi\pi}$\,$\approx$ 0.19\,keV. Accepting that $2m_\pi<
$\,$\sqrt{s}<$\,0.8\,GeV is the region of the wide $\sigma(600)$
resonance, we obtain $\langle\Gamma_{\sigma\to\gamma\gamma}
\rangle_{\pi\pi}$\,$\approx$\,0.45\,keV by analogy with Eq. (8).

Note that the contributions from the $\omega(782)$ and $h_1(1170)$
exchanges to the $S$-wave amplitude of the
$\gamma\gamma$\,$\to$\,$\pi^0\pi^0$ reaction have opposite signs and
cancel each other.\\

This work was supported in part by the RFFI Grant No. 07-02-00093
from the Russian Foundation for Basic Research and by the
Presidential Grant No. NSh-1027.2008.2 for Leading Scientific
Schools.


\begin{thebibliography}{99}
\bibitem{fn17} We verify that, if the approximation includes the
               contribution from the $D_{\lambda=0}$ wave; i.e., if the expression
               $|a+c\,d^2_{00}(\theta)|^2+|b\,d^2_{20} (\theta)|^2$ with four free
               parameters is used to describe the angular distributions, it is
               impossible to reliably determine the absolute value and relative
               phase of the amplitude $c$, because this expression does not provide
               any noticeable improvement to the approximation. Note that the
               so-called SD and SDG approximations used in [5] lead to the choice
               of the nonphysical values of the parameters and, as a result, to
               negative differential cross sections (see Fig. 7 in [5]). For this
               reason, they cannot provide a reliable tool for the partial wave
               analysis.
\end{thebibliography}
\end{document}